\documentclass[aps,prb,floatfix,twocolumn,nobibnotes,longbibliography,superscriptaddress]{revtex4-2}
\usepackage{graphicx}
\usepackage{todonotes}
\usepackage{amsmath} 
\usepackage{amsmath}
\usepackage{upgreek}
\usepackage{color}
\usepackage{standalone}
\usepackage{pgfplots}
\pgfplotsset{compat=1.5}
\usetikzlibrary{positioning}
\usepackage{array}
\newcolumntype{C}[1]{>{\centering\let\newline\\\arraybackslash\hspace{0pt}}m{#1}}
\usepackage{multirow}
\usepackage{hyperref}
\usepackage{float}
\usepackage{todonotes} 
\usepackage{rotating}
\usepackage{gensymb}\usepackage{xcolor}

\usepackage{lmodern}        
\begin{document}

\newcommand{\dd}[1]{\frac{\text{d}}{\text{d}#1}}
\newcommand{\de}{\text{d}}
\newcommand{\chipe}{ \chi^m_{\perp} }
\newcommand{\chipa}{ \chi^m_{\parallel} }
\renewcommand{\phi}{ \varphi }
\newcommand{\unitvec}[1]{\mathbf{\hat #1}}
\newcommand{\ve}[1]{{\mathbf{#1}}}
\newcommand{\degC}[1]{{$^{\rm\circ}$}}


	\title{Paraelectric KH$_2$PO$_4$ Nanocrystals in Monolithic Mesoporous Silica:\\ Structure and Lattice Dynamics} 

\author{Yaroslav Shchur }
\affiliation{Institute for Condensed Matter Physics, 1 Svientsitskii str., 79011, Lviv, Ukraine}
\email{shchur@icmp.lviv.ua}
\author{Andriy V. Kityk}
\affiliation{Faculty of the Electrical Engineering, Czestochowa University of Technology, Al. Armii
Krajowej 17, 42-200, Czestochowa, Poland}
\author{Viktor V. Strelchuk}
\affiliation{V.E. Lashkaryov Institute of Semiconductor Physics, National Academy of Sciences of Ukraine, 41 Prosp. Nauky, 03028 Kyiv, Ukraine}
\author{Andrii S. Nikolenko}
\affiliation{V.E. Lashkaryov Institute of Semiconductor Physics, National Academy of Sciences of Ukraine, 41 Prosp. Nauky, 03028 Kyiv, Ukraine}
\author{Nazariy A. Andrushchak}
\affiliation{Lviv National Polytechnic University, 12 S. Bandery str., 79013, Lviv, Ukraine}
\author{Patrick Huber}
\affiliation{Hamburg University of Technology, Center for Integrated Multiscale Materials Systems CIMMS, Ei\ss endorferstr. 42, 21073 Hamburg, Germany}
\affiliation{Center for X-ray and Nano Sciences CXNS, Deutsches Elektronen-Synchrotron DESY, 22603 Hamburg, Germany}
\affiliation{Centre for Hybrid Nanostructures CHyN, Hamburg University, 22607 Hamburg, Germany}
\author{Anatolii S. Andrushchak}
\affiliation{Lviv National Polytechnic University, 12 S. Bandery str., 79013, Lviv, Ukraine}

	
\begin{abstract} 
Combining dielectric crystals with mesoporous solids allows a versatile design of functional nanomaterials, where the porous host provides a mechanical rigid scaffold structure and the molecular filling adds the functionalization. Here, we report a study of the complex lattice dynamics of a SiO$_2$:KH$_2$PO$_4$ nanocomposite consisting of a monolithic, mesoporous silica glass host with KH$_2$PO$_4$ nanocrystals embedded in its tubular channels $\sim$12 nm across. A micro-Raman investigation performed in the spectral range of 70-1600 cm$^{-1}$ reveals the complex lattice dynamics of the confined crystals. Their Raman spectrum resembles the one taken from bulk KH$_2$PO$_4$ crystals and thus, along with X-ray diffraction experiments, corroborates the successful solution-based synthesis of KH$_2$PO$_4$ nanocrystals with a structure analogous to the bulk material. We succeeded in observing not only the high-frequency internal modes ($\sim$900-1200 cm$^{-1}$), typical of internal vibrations of the PO$_4$ tetrahedra, but, more importantly, also the lowest frequency modes typical of bulk KH$_2$PO$_4$ crystals. The experimental Raman spectrum was interpreted with a group theory analysis and  first-principle lattice dynamics calculations. The calculated phonon spectrum reveals a lattice instability against two optical modes probably related to an intrinsic anharmonicity of the vibrational dynamics of KH$_2$PO$_4$ crystals originating from the uncertainty of proton localization within the network of hydrogen bonds. The analysis of calculated \textit{eigen}-vectors indicates the involvement of hydrogen atoms in most phonon modes corroborating the substantial significance of the hydrogen subsystem in the lattice dynamics of paraelectric bulk and of KH$_2$PO$_4$ crystals in extreme spatial confinement. A marginal redistribution of relative Raman intensities of the confined compared to unconfined crystals presumably originates in slightly changed crystal fields and interatomic interactions, in particular for the parts of the nanocrystals in close proximity to the silica pore surfaces.

keywords: Raman scattering, KH$_2$PO$_4$, porous SiO$_2$, phonon, density functional theory, confinement effect

\end{abstract}

\maketitle

\section{Introduction}
In solid state physics there are several compounds which have been used as benchmark materials for testing fundamental theoretical and experimental concepts. Typical examples are the perovskite family, in particular BaTiO$_3$ crystals with a phase transition of the displacive type \cite{Rabe,Scott}, Rochelle salt KNaC$_4$H$_4$O$_6\cdot$4H$_2$O as a famous piezoelectric crystal \cite{Jona} and LiNbO$_3$ as a prominent optical, acousto- and non-linear optical compound \cite{Rabe,Kitamura}. 

\begin{figure}[]
\begin{center}
 \includegraphics [width=.9\columnwidth] {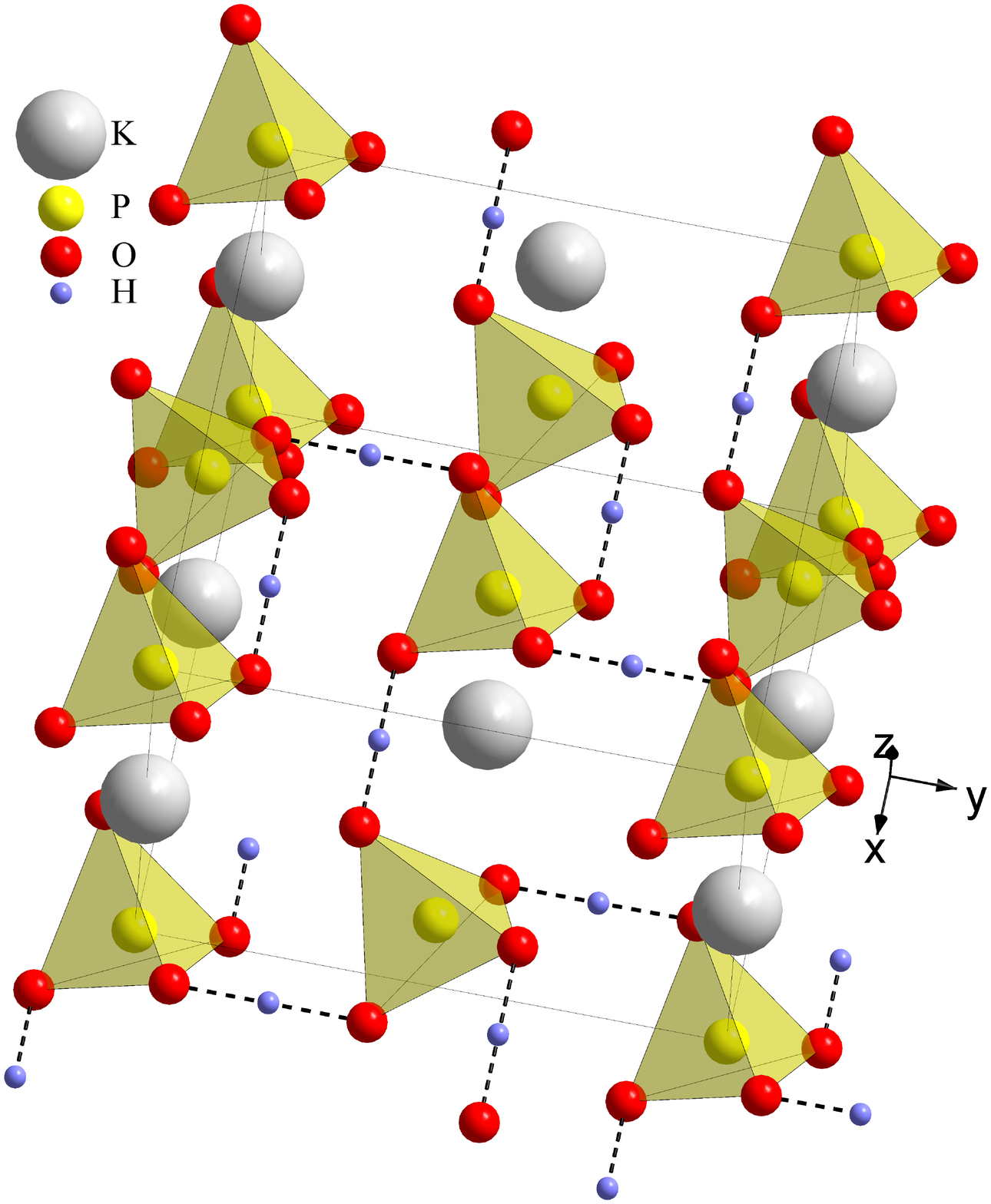}
\end{center}
\caption[]{Crystal structure of KH$_2$PO$_4$ crystals at room temperature (space group I$\bar{4}$2d) according to Ref. \cite{Itoh1998}.} Hydrogen bonds are indicated by dashed lines. \label{struct}
\end{figure}

An especially prominent role plays here also the extended family of hydrogen-bonded crystals of KH$_2$PO$_4$ type \cite{Jona,Lines}. These crystals cover a wide application range in quantum electronics, e.g. as optical modulators, frequency multiplicators in the optical range and Q-switchers in pulsed lasers. Another no less significant feature of these crystals is related to the physics of structural phase transitions. Since the phase transition in KH$_2$PO$_4$ from the room temperature I$\bar{4}$2d (No.122) phase to the low-temperature Fdd2 (No.43) phase \cite{Bacon,Itoh1998,Miyoshi2011} is accompanied with proton ordering in hydrogen bonds O-H$\cdot\cdot\cdot$O (see crystal structure in Figure \ref{struct}), the KH$_2$PO$_4$ crystal was widely treated as a benchmark compound for the study of ``order-disorder" structural phase transitions. There is a huge scientific literature devoted to the nature of structural ferroelectric phase transition in KH$_2$PO$_4$. Even cursory analysis of theoretical models and experimental studies aimed at interpreting the phase transition mechanism in KH$_2$PO$_4$ goes far beyond the scope of our research. We may only mention here the most significant, milestone ideas played important role in proper understanding the mechanism of ferroelectric phase transition in KH$_2$PO$_4$. The first microscopic model suggested in 1941 by Slater \cite{Slater} accounts various arrangements of four protons around the selected phosphate PO$_4$ group. This model has become a basis for so-called "proton ordering model" which was further elaborated by R. Blinc in 1960 \cite{Blinc1960} who proposed an idea of proton tunnelling in {symmetric double-well potential} on hydrogen bond. This "proton ordering model" had an enormous impact on the entire structural phase transitions theory in solid state physics. After the series of experimental studies \cite{Tominaga1,Nelmes1980}, it turned out that the phase transition mechanism in KH$_2$PO$_4$ is a more complicated one and cannot be described as solely of "order-disorder" type. It is rather of mixed kind, i.e. it contains "order-disorder" and displacive contributions. Proton ordering in O-H$\cdot\cdot\cdot$O hydrogen bonds is inseparably linked to the PO$_4$ tetrahedra distortion and to the displacements of K atoms and of the PO$_4$ groups in \textit{z}-direction. Therefore the "proton ordering model" was modified by Kobayashi in 1968 \cite{Kobayashi1968} who proposed to account the bonded proton-lattice vibrations. Since the IR and Raman scattering data revealed the structural disorder of PO$_4$ tetrahedral groups even in paraelectric phase, Blinc and Zeks expended the Kobayashi's approach to the model ascribing the spin 1/2 to the whole H$_2$PO$_4$ group which is coupled to K motion \cite{Blinc1982}. All further microscopic models were based, to a greater or lesser extent, on the  elements of formerly suggested models.

We would like to point here out the particular significance of Raman scattering studies in proper understanding the peculiarities of structural phase transition in KH$_2$PO$_4$. Raman scattering distinguishes the components of molecular polarizability tensor which are related to site symmetry of selected crystal constituents and to the whole crystal symmetry. Owing to the series of Raman scattering experiments \cite{Tominaga1,Tominaga} there were unambiguously concluded that the momentary local symmetry of PO$_4$ groups in paraelectric phase is lower than \textit{$\bar{4}$} predicted for I$\bar{4}$2d space group at room temperature and should coincide with the site symmetry \textit{2} inherent for PO$_4$ groups in ferroelectric Fdd2 phase. In other words, owing to the dynamical disorder of PO$_4$ groups in paraelectric phase the \textit{$\bar{4}$} site symmetry is the time and space averaged symmetry between two marginal site positions of \textit{2} local symmetry.

The possibilities of the wide isomorphic substitution of K atoms by Li, Cs, Rb, Tl ones and by NH$_4$ groups and the deuteration, i.e. the H and D substitution, gives another very sensitive tool for a detailed investigation of the peculiarities of the phase transition mechanism in KH$_2$PO$_4$ crystals. Owing to such rich physical and structural features the KH$_2$PO$_4$ crystals and more generally the large family of KH$_2$PO$_4$ compounds have become the target materials for probing the reliability of a vast variety of scientific concepts and experimental methods since the early sixties of the last century.

This work is involved in the broader scientific and technological context to embed dielectric crystals in monolithic nanoporous hosts to design new hybrid materials owing their optical functionalization from the molecular pore filling and their mechanical rigidity from the optically transparent, porous glass scaffold \cite{Huber2020}. To this end, it is important to achieve a detailed proof of the proper formation of the dielectric crystals in the nanoporous host as well as a rigorous understanding of the structural and dynamical properties of the extremely confined crystals compared to the unconfined, bulk material.

Specifically, in this work we aim to investigate the crystal structure of a nanocomposite based on KH$_2$PO$_4$ nanocrystals deposited by solvent evaporation in tubular nanopores of SiO$_2$ matrix by Raman scattering and X-ray diffraction. The main questions to be addressed are: Do we obtain nanocrystals of KH$_2$PO$_4$ compound inside the pore space? Do the main structural features of KH$_2$PO$_4$ crystal preserve in confined nanocrystals? How can the experimental Raman scattering data be related to \textit{ab-initio} calculations of the phonon spectrum of KH$_2$PO$_4$ crystals in its room temperature phase.

Note that a small, preliminary part of our experimental results have been reported in conference proceedings \cite{ICTON}. Nonlinear optical study on KH$_2$PO$_4$ embedded into nanoporous anodized alumina oxide matrix has been presented in the Ref. \cite{Andr1}.

\section{Experimental and computational details}
A monolithic mesoporous silica SiO$_2$ membrane of 300 $\mu$m thickness was used as a host matrix for the fabrication of a SiO$_2$:KH$_2$PO$_4$ nanocomposite. The porous SiO$_2$ was prepared by thermal oxidation of mesoporous silicon, pSi, at 1073~K for 12~h. The pSi membranes were synthesized by electrochemical anodic etching of highly p-doped (100)-oriented silicon wafers employing a mixture of concentrated fluoric acid and ethanol (volume ratio 2:3) as electrolyte. The resulting pSiO$_2$ membranes consist of an array of parallel-aligned nanochannels of mean diameter D=12.0$\pm$0.5 nm and exhibits a porosity P=40$\pm$2 \%, as determined by recording a volumetric nitrogen sorption isotherm at 77~K.

The KH$_2$PO$_4$ nanocrystals have been embedded into the nanopores of the SiO$_2$ matrix by imbibing a saturated water-KH$_2$PO$_4$ solution into the mesoporous silica \cite{Gruener_2011} and subsequent spontaneous evaporation of water. This results in the formation of nanocrystals inside the nanochannels. Note that a certain excess of KH$_2$PO$_4$ usually remains on the SiO$_2$ membrane surface forming bulk polycrystalline textural layers, often of (100)-orientation. This layer was carefully removed by a razor. The sample quality was tested by X-ray diffraction (XRD) and scanning electron microscopy. Only the samples with flat and comparatively uniform surfaces were selected for spectroscopic investigations. The XRD measurements ($\theta$/2$\theta$ - scans) have been performed using commercial X-ray diffractometer DRON-3 (Cu-anode, K$\alpha$  radiation, $\lambda$=1.5406 $\AA$).

Micro-Raman measurements were performed at room temperature in backscattering geometry using a 785~nm laser as a light source and a micro-Raman spectrometer Horiba Jobin Yvon T64000 equipped with an Olympus BX41 microscope and an electrically-cooled Si CCD detector. The exciting radiation was focused on the sample surface with an $\times$50/NA 0.75 optical objective, giving a laser spot diameter of ~1 $\mu$m. The laser power on the sample surface was varied in the range of 1-2~mW. Raman spectra were registered at different points of the sample and from both sides of the sample.

Calculation of phonon spectrum was carried out within the density functional perturbation approach using the generalized gradient approximation (GGA) \cite{Baroni} in Perdew-Burke-Ernzerhof parameterization \cite{Perdew} while taking into consideration the long-range dispersion effects as suggested by Grimme \textit{et al.} (DFT-D3 method) \cite{Grimme}. The \textit{ab-initio} package ABINIT \cite{Abinit1,Abinit2} was used in our work. The norm-conserving ONCVPSP pseudopotentials were utilised with K(3s$^2$3p$^6$4s$^1$), P(3s$^2$3p$^3$), O(2s$^2$2p$^4$) and H(1s$^1$) valence states. The convergence analysis was performed concerning both the sampling of the Brillouin zone using the Monkhorst-Pack scheme \cite{Monkhorst} and the kinetic energy cut-off for plane-wave calculations.

We used experimental crystal structure resolved at room temperature in paper \cite{Itoh1998}. Both the lattice parameters and atom coordinates were relaxed during structural optimization which was done within the Broyden-Fletcher-Goldfarb-Shanno algorithm \cite{Broyden}. The restriction of preserving the I$\bar{4}$2d space group was imposed during structural optimization. Such symmetry restriction always keeps all protons in the centers of O-H$\cdot\cdot\cdot$O hydrogen bonds. The maximal forces acting on each atom were reduced to the value lower than 1.9$\times$10$^{-8}$ eV/$\r{A}$. The 4$\times$4$\times$4 grid for Brillouin zone sampling was used and the cut-off energy, E$_{cut}$, was set to 1632 eV with a cut-off smearing of 13.6 eV.

\section{Results}
\subsection{X-ray diffraction}
\begin{figure}[!ht]
\begin{center}
 \includegraphics [width=0.9\columnwidth] {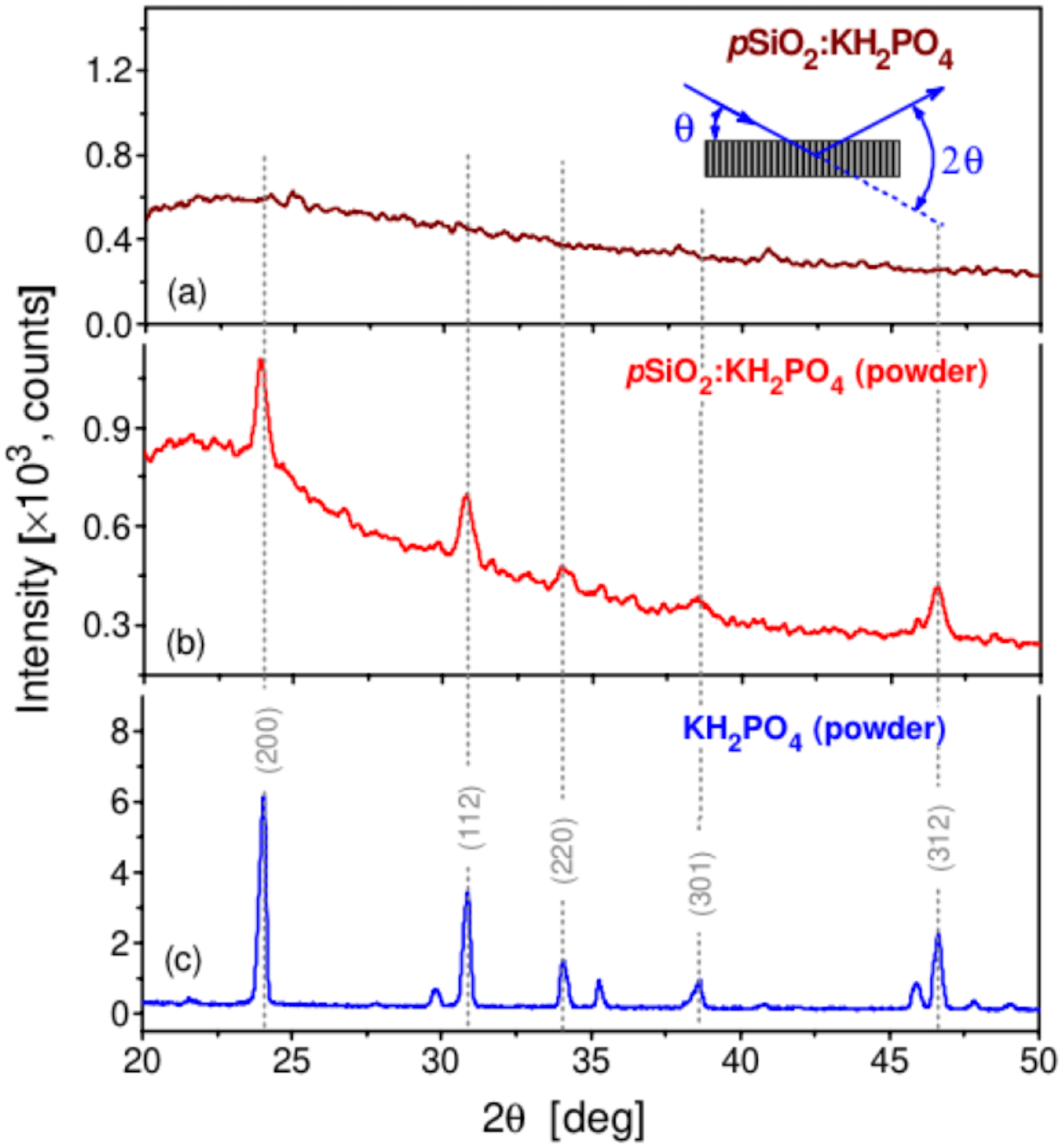}
\end{center}
\caption[]{X-ray diffraction patterns ($\theta$-2$\theta$ scans) recorded on a monolithic membrane of a SiO$_2$:KH$_2$PO$_4$ nanocomposite (a), its powder counterpart (b) and a powder KH$_2$PO$_4$ sample as reference (c).} \label{xrd}
\end{figure}

In Figure \ref{xrd} we show X-ray diffraction (XRD) $\theta$-2$\theta$ scans recorded from monolithic membranes of the SiO$_2$:KH$_2$PO$_4$ nanocomposite (Figure \ref{xrd}(a)), its powder counterpart (Figure \ref{xrd}(b)) prepared by cruching a nanocomposite monolith, and a powder of bulk KH$_2$PO$_4$ crystallites as a reference sample (Figure \ref{xrd}(c)). The powder XRD pattern of SiO$_2$:KH$_2$PO$_4$ composite well reproduces most intensive Bragg reflection peaks typical of the bulk reference KH$_2$PO$_4$ pattern confirming thus a presence of KH$_2$PO$_4$ nanocrystals in the silica nanochannels. An analysis of the crystallite size based on the widths of the detected Bragg peaks analyzed with the Scherrer ansatz gives a value of 21.4$\pm$1 ~nm. This indicates that the crystallites grow along the tubular pore axis slightly larger than perpendicular to it, where they are limited in size by the pore diameter. Note that the strong scattering background results from the amorphous glass matrix with a broad intensity maximum at about 21 deg typical of the short range order in amorphous silica \cite{Huber1999}.

An absence of any peaks in the XRD diffractogram of the monolithic, uncrushed sample, on the other hand, suggests that a very pronounced preferred orientation of the nanocrystals inside the silica nanochannels is present. However, since in our scattering experiment the wave vector transfer can be only kept parallel to the monolith's surface normal, we can not characterize the texture more precisely. Note that due to the specific tetrahedral structure of KH$_2$PO$_4$ crystals the (001)-atomic planes do not give a Bragg reflection. Thus, if the prominent $<$001$>$ orientation is growing parallel to the long pore axis of the channels, this would explain the absence of any reflections in the XRD pattern of the monolithic sample for our scattering geometry. Such a preferred crystal growth in nanochannels is known from many other compounds, e.g. nitrogen, hydrocarbons and liquid crystals. It can often be traced to the crystallization kinetics, e.g. Bridgman type crystal growth in anisotropic confinement \cite{Henschel2009,Huber2015}. Detailed textural insights on the KDP crystallization in pore space would only be achievable by 3-D reciprocal space mapping, as it can be performed in a versatile manner on monolithic mesoporous membranes at synchrotron-based X-ray sources \cite{Sentker2018, Sentker2019}.

\subsection{Raman scattering}
\begin{figure}[!ht]
\begin{center}
\includegraphics[width=1\columnwidth]{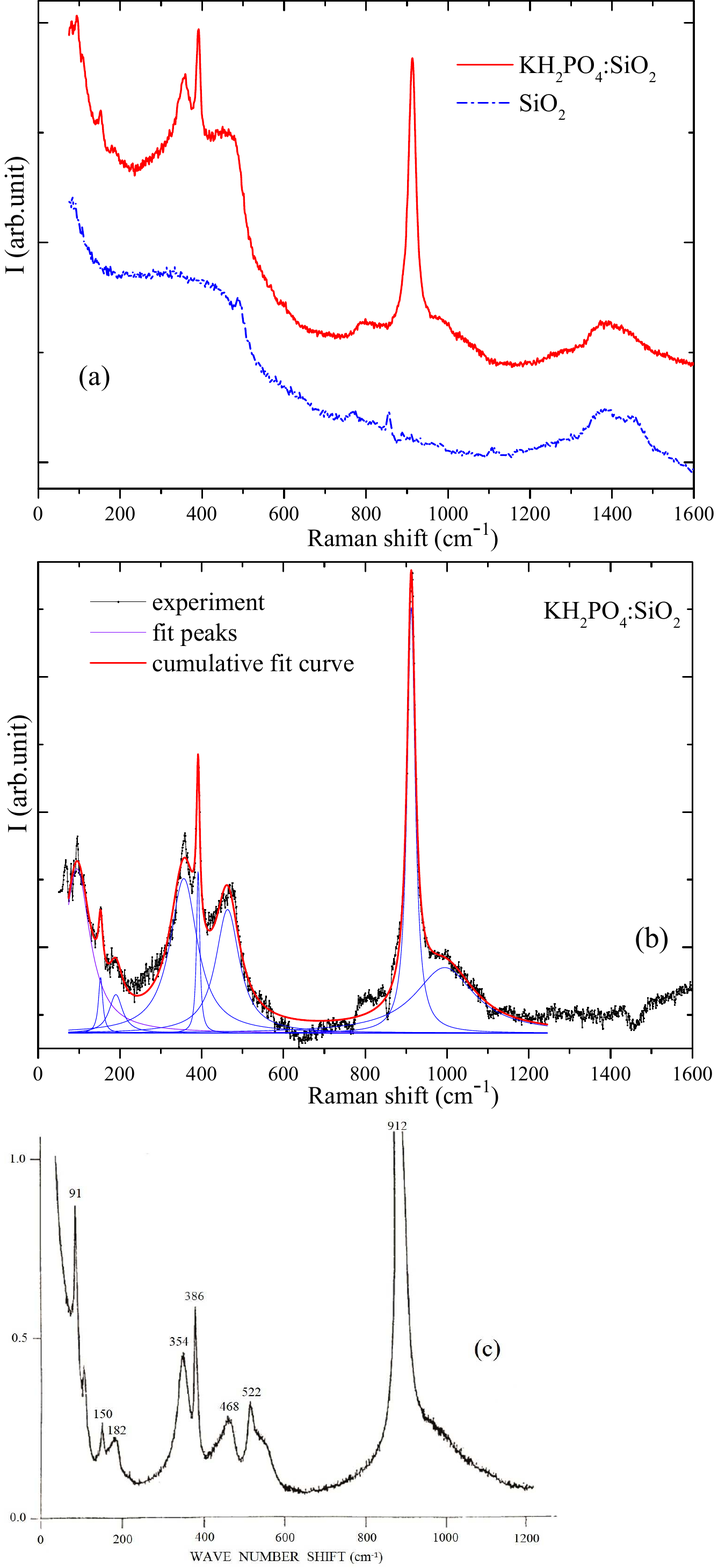}
\end{center}
\caption[]{(a) Raman spectra of SiO$_2$:KH$_2$PO$_4$ nanocomposite and of the host SiO$_2$ matrix; (b) result of a subtraction of Raman data of SiO$_2$:KH$_2$PO$_4$ nanocomposite and that of the host SiO$_2$ matrix. Blue lines correspond to Lorentzian constituents, red line to a cumulative fitting curve; (c) Raman spectrum of bulk KH$_2$PO$_4$ reproduced from Ref.~ \cite{Kim}. All spectra were measured at room temperature.} \label{Raman}
\end{figure}

Raman spectra recorded at room temperature from the SiO$_2$:KH$_2$PO$_4$ nanocomposite and from the host SiO$_2$ matrix are presented in Figure \ref{Raman}(a). We measured identical spectra at different places of the sample from both sides indicating a good reproducibility. The result of a subtraction of the two Raman spectra depicted in Figure \ref{Raman}(a) is shown in Figure \ref{Raman}(b). It should be representative of the newly synthesized material in pore space. In Figure \ref{Raman}(c) we also present a Raman spectrum of a bulk KH$_2$PO$_4$ crystal, as reproduced from Ref. \cite{Kim}. Comparing Figures \ref{Raman}(b) and \ref{Raman}(c) one concludes that the subtracted Raman spectrum agrees in all main features with the spectrum taken from bulk KH$_2$PO$_4$\cite{Kim}.

The interpretation of the KH$_2$PO$_4$ Raman spectra can be based on a group theory analysis and the vibrational peculiarities of structural constituents of KH$_2$PO$_4$. At room temperature, KH$_2$PO$_4$ crystallizes in tetragonal symmetry with I$\bar{4}$2d space group (Z=4). 48 normal modes of the primitive lattice are classified in the Brillouin zone centre as follows
\\
\centerline{$\Gamma$=4A$_1$(R)+5A$_2$(inact.)+6B$_1$(R)+7B$_2$(R,IR)+13E(R,IR),}
where R and IR indicate Raman and infrared activity, respectively. Modes with A$_2$ symmetry are neither Raman nor IR active. Since the irreducible representation E is two-dimensional, all modes of E symmetry are twice degenerated. Performing the \textit{eigen}-vector analysis in the spirit of Maradundin and Vosko \cite{Maradudin} and using the correlation diagram for internal modes of PO$_4$ tetrahedra of KH$_2$PO$_4$ crystals in the I$\bar{4}$2d phase \cite{Tominaga}, one may perform a tentative assignment of phonon modes to the different spectral ranges (see Table \ref{Classif}). Making such a classification we took into consideration some basic statements, i.e. normally, external (lattice) vibrations of K and PO$_4$ (or H$_2$PO$_4$) groups are placed below $\sim$250 cm$^{-1}$, internal vibrations of tetrahedral PO$_4$ groups are distributed over the bending, $\nu_2$ and $\nu_4$, and stretching, $\nu_1$ and $\nu_3$, vibrations. As was shown based on lattice dynamics calculations, owing to strong covalent phosphorous-oxygen bonding the frequencies of these internal PO$_4$ vibrations insignificantly depend on the crystal environment in all compounds of KH$_2$PO$_4$-type, namely CsH$_2$PO$_4$, RbH$_2$PO$_4$, TlH$_2$PO$_4$, PbHPO$_4$ and their deuterated analogues \cite{CDP2,DRDP,TDP,LHP2}. In aqueous solution, the following frequencies of the free PO$_4$ tetrahedra were detected, $\nu_2$=420 cm$^{-1}$, $\nu_4$= 560 cm$^{-1}$, $\nu_1$= 940 cm$^{-1}$, $\nu_3$= 1020 cm$^{-1}$ (Ref. \cite{Nakamoto}).

\begin{figure}[ht]
\begin{center}
\includegraphics[width=0.7\columnwidth]{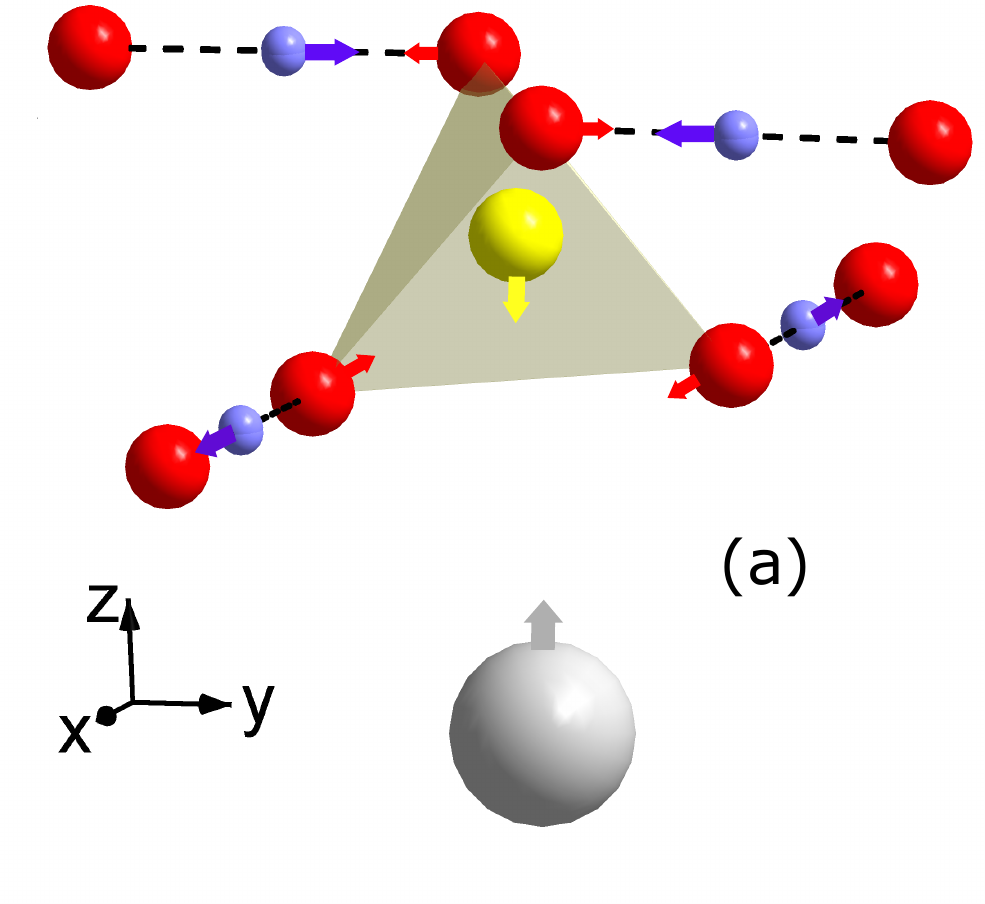}\\
\includegraphics[width=0.7\columnwidth]{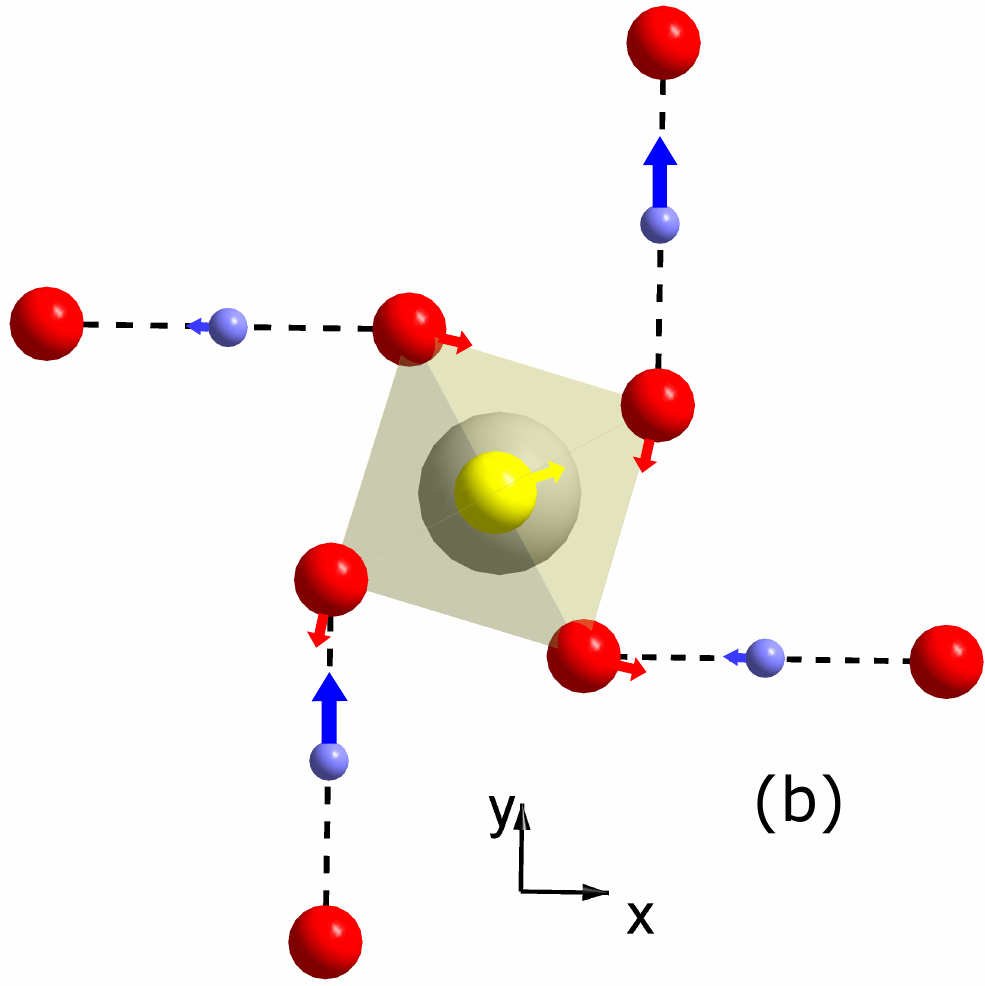} \\ \includegraphics[width=0.7\columnwidth]{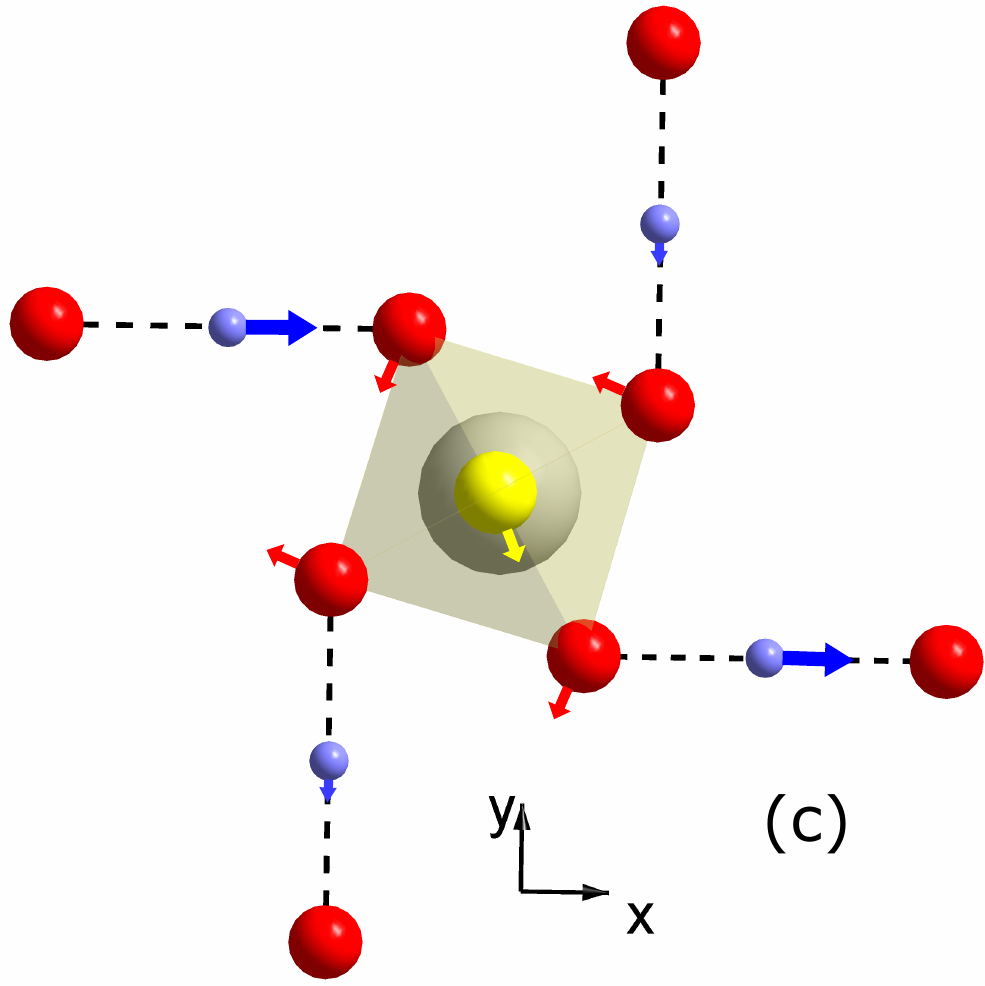}
\end{center}
\caption[]{Simplified illustration of unstable optic phonon modes. (a) ferroactive B$_2$ mode with imaginary frequency \textit{Im} 539 cm$^{-1}$; (b) two-fold degenerated E mode with imaginary frequency \textit{Im} 514 cm$^{-1}$. The color presentation of the constituent atoms is the same as in Figure \ref{struct}.} \label{eigen}
\end{figure}

Therefore, the internal PO$_4$ vibrations are a reason of two comparatively separated spectral ranges within the $\sim$300-600 cm$^{-1}$ and $\sim$900-1100 cm$^{-1}$ regions. On the first glance, the proton vibrations in $O-H$$\cdot\cdot\cdot$$O$ hydrogen bonds should be located in the highest frequency range, above $\sim$1200 cm$^{-1}$, and this is true for many highest energy lines in the Raman spectrum. However, it turned out that the real phonon dynamics of a number of hydrogen-bonded crystals is much more complicated and some hydrogen vibrations are mixed with internal PO$_4$ group vibrations or even with low-frequency lattice vibrations. As we already mentioned in the Introduction, the nature of proton dynamics in KH$_2$PO$_4$ type crystals has been debated for many decades and this topic goes far beyond the main purpose of our paper.

Summarizing our symmetry consideration one may construct the general classification of KH$_2$PO$_4$ crystal phonon modes at room temperature  (paraelectric phase, see Table \ref{Classif}). According to this Table, 9, 7, 5 and 7 different Raman modes may be expected as observable in the lowest frequency range below $\sim$300 cm$^{-1}$, mid-frequency $\sim$300-600 cm$^{-1}$, high-frequency $\sim$900-1100 cm$^{-1}$ and highest-frequency $>$1200 cm$^{-1}$ regions, respectively. Although, the total number of experimentally observed Raman modes may be even larger owing to the lowering of the ideal crystal symmetry in real imperfect crystals and the degeneracy lifting of doubly degenerated phonon modes of E symmetry.

\begin{table*} []
\caption{Normal mode classification of KH$_2$PO$_4$ in the paraelectric I$\bar{4}$2d phase. The polarization of proton modes is indicated in the cartesian coordinate system x$\parallel$a, y$\parallel$b, z$\parallel$c.}
 \label{Classif}
\footnotesize
\begin{tabular}{lcccc}
\\\hline
&translations&internal&internal& proton\\
symmetry&+&bending&stretching&modes\\
&librations&PO$_4$ modes&PO$_4$ modes&\\
\hline
&$<$300 cm$^{-1}$&300-600 cm$^{-1}$&900-1100 cm$^{-1}$&$>$1200 cm$^{-1}$ \\
 \hline
4A$_1$(R)&1 libr.&$\nu_2$&$\nu_1$&1:~H(x)\\
5A$_2$(inact.)&1 libr.&$\nu_2$&$\nu_1$&2:~H(y,z)\\
6B$_1$(R)&2 ext.&$\nu_2$,$\nu_4$&$\nu_3$&1:~~H(x)\\
7B$_2$(R,IR)&1~acoust.+1 ext.&$\nu_2$,$\nu_4$&$\nu_3$&2:~H(y,z)\\
13E(R,IR)&1~acoust.+2 libr.+3 ext.&2$\nu_4$&2$\nu_3$&3:~~H(x,y,z)\\
\hline
$\sum$ Raman&9&7&5&7\\
\hline
\end{tabular}
\end{table*}

The experimental Raman spectra were fitted by a series of Lorentzian profiles. To that end we introduced eight Lorentzian constituents and reached an acceptable representation of the experimental Raman spectrum by the cumulative fitting curve, see Figure \ref{Raman}(b). The experimental Raman frequencies encompassing three lowest frequency modes 95, 153 and 190 cm$^{-1}$ which correspond to external lattice vibrations, three mid-frequency modes 356, 391 and 463 cm$^{-1}$ and two high-frequency modes 912 and 994 cm$^{-1}$ are listed in Table \ref{Compar}.

These modes taken from the SiO$_2$:KH$_2$PO$_4$ nanocomposite correspond to 91, 150, 182, 354, 386, 468, 522 and 912 cm$^{-1}$ Raman modes of the bulk KH$_2$PO$_4$ crystal (see Figure \ref{Raman}(c)) previously published in Ref.~\cite{Kim}. There is a very good agreement between Raman frequencies of our SiO$_2$:KH$_2$PO$_4$ nanocomposite and those presented in Ref.~\cite{Kim} with exception of only one narrow line at 522 cm$^{-1}$. This mode, present in bulk KH$_2$PO$_4$ Raman spectrum, has no immediate counterpart in the SiO$_2$:KH$_2$PO$_4$ Raman spectrum. However, accounting for the rather large full-width-half-maximum of this line, $>$ 70 cm$^{-1}$, one may suppose that the counterpart of the 522 cm$^{-1}$ line overlaps with the 463 cm$^{-1}$ line observed in the nanocomposite Raman spectrum and cannot be detected in spectrum owing to the low intensity of this mode. There appears a question why the intensity of this vibrational mode becomes much lower comparing with corresponding mode in Raman spectrum of bulk crystal. Probably, this is a result of impact of confinement effect on vibrational spectrum of KH$_2$PO$_4$ nanocrystals introduced in SiO$_2$ nanopores.

The most intensive line at the frequency 912 cm$^{-1}$ observed in both spectra most probably corresponds to the stretching $\nu_1$ mode. The correlation between the bending internal vibrations, $\nu_2$ and $\nu_4$, observed in the mid-frequency regions of both the nanocomposite and the bulk material indicates the presence of covalently bonded tetrahedral PO$_4$ groups in our nanocomposite compound. Although the existence of PO$_4$ groups itself cannot undoubtedly prove the KH$_2$PO$_4$ crystal structure of the newly synthesized compound, because these PO$_4$ groups may be involved in some different structure, not necessarily in the KH$_2$PO$_4$ one. The most valuable confirmation of the presence of the KH$_2$PO$_4$ crystal structure in our newly synthesized confined crystals are the three lowest frequency modes placed at 95, 153 and 190 cm$^{-1}$ (see Figure \ref{Raman}(b)). These modes correspond to external vibrations of K$^-$ and PO$_4^{-3}$ ions and can be treated as direct markers of the KH$_2$PO$_4$ crystal structure. However, comparing the Raman spectra of our SiO$_2$:KH$_2$PO$_4$ sample (Figure \ref{Raman}(b)) and bulk KH$_2$PO$_4$ (Figure \ref{Raman}(c)) one may reveal that the relative intensities of three low-frequency phonon modes differ from those counterparts of the bulk KH$_2$PO$_4$. The lowest frequency mode at 91 cm$^{-1}$ is the most intensive mode in the bulk KH$_2$PO$_4$ spectrum below $\sim$800 cm$^{-1}$, whereas the most intensive line in SiO$_2$:KH$_2$PO$_4$ spectrum below $\sim$800 cm$^{-1}$ is the internal bending PO$_4$ vibration at 391 cm$^{-1}$. The similar redistribution of line intensities observed in nanostructured and bulk forms of KH$_2$PO$_4$ concerns to the most intensive PO$_4$ stretching $\nu_1$ mode at 912 cm$^{-1}$. As one may judge from Figure \ref{Raman}(c), the intensity of the mode at 912 cm$^{-1}$ is more than twice as large compared with any other mode in the spectrum, whereas the intensity of this stretching $\nu_1$ mode in the spectrum of nanocomposite SiO$_2$:KH$_2$PO$_4$ sample (Figure \ref{Raman}(b)) is only $\sim$1.6 times larger than the intensity of the mode at 391 cm$^{-1}$. A similar argument concerning the intensity mode redistribution may explain why we do not observe in SiO$_2$:KH$_2$PO$_4$ spectrum the mode near 522 cm$^{-1}$ detected in bulk form. The reason may be too low intensity of such a mode which hinders its experimental detection. Based on these experimental observations one may conclude that the Raman spectrum of the newly synthesized compound corresponds in the main features with the vibrational spectrum of bulk KH$_2$PO$_4$ crystal in the room temperature I$\bar{4}$2d phase. The confinement effect has a certain but not crucial impact on the nanocrystals synthesized within the $\sim$12 nm nanopores. Probably it changes slightly the internal crystal field and is partly modifying interatomic interactions, in particular for those parts of the nanocrystals in close proximity of the external electrical fields of the polar OH-terminated silica pore surfaces {similarly to the case of OH-decorated silicon pore surfaces \cite{porSi}.} It is worth noting that a similar effect of some intensity redistribution of Raman lines was recently observed in spectra of few-layer and monolayer (thickness 0.6 nm) thin-film samples of TaS$_2$ crystals when compared to that of the bulk crystal \cite{TaS2}.

{Summarizing this section one may state that the interaction between the host SiO$_2$ matrix and nanosized KH$_2$PO$_4$ crystal is governed through the pure spatial confinement effect as well as the interface interaction of two media, SiO$_2$ and KH$_2$PO$_4$. Applying solely Raman scattering technique one is not able to disentangle these two effects. However, as follows from our data the overall impact is not dominant since the main features of Raman spectra taken from bulk and nanostructured KH$_2$PO$_4$ are similar.}

\section{\textit{Ab initio} calculation and discussion}
To get a deeper insight into our experimental spectroscopic data we have undertaken \textit{ab-initio} lattice dynamics calculations. The theoretical treatment of lattice dynamics of KH$_2$PO$_4$ type crystals has its challenges due to the dynamic disorder of both H atoms in the O-H$\cdot\cdot\cdot$O hydrogen bonds and PO$_4$ tetrahedral groups. It is known that, at room temperature, H atoms in  KH$_2$PO$_4$ occupy with equal probability two off-center positions in hydrogen bonds. A structural transition to the low-temperature Fdd2 phase is accompanied by a proton ordering in one of two possible positions in the O-H$\cdot\cdot\cdot$O bonds and correlates with displacements of K and P atoms and distortions of PO$_4$ groups. To preserve the high-temperature I$\bar{4}$2d symmetry during the lattice dynamics calculations one needs to place all hydrogen atoms in the centre of each hydrogen bond, even though they occupy with equal probability two off-center positions in the O-H$\cdot\cdot\cdot$O bonds (split position of hydrogen atoms). This indispensable model approximation is widely used in many atomistic and \textit{ab initio} considerations of lattice dynamics of hydrogen bonded crystals \cite{CDP2,Zhang}. Unfortunately, adopting this symmetry trick we \textit{a priori} weaken our theoretical description and create a source of disagreement between experimental and theoretical vibrational frequencies of  KH$_2$PO$_4$ crystals. It becomes obvious from the comparison of experimental Raman and theoretical phonon frequencies calculated in the Brillouin zone center (see Table \ref{Compar}), where all frequencies of Raman active modes are classified according to irreducible representations of the I$\bar{4}$2d group. As can be seen from this table, there are two modes of B$_2$ and E symmetry with imaginary frequencies, \textit{Im} 539 and \textit{Im} 514 cm$^{-1}$, respectively. In lattice dynamics theory, an imaginary frequency evidences a lattice instability. Note that the same lattice instability against two phonon modes was recently detected in similar \textit{ab initio} calculations of a phonon spectrum of KH$_2$PO$_4$ crystal \cite{Menchon}. Since our theoretical consideration is performed in harmonic approximation, the resulting lattice instability implies that the significant anharmonic lattice contribution must balance the negative harmonic part to stabilize the phonon dynamics of a real crystal. However, a proper accounting of lattice anharmonicity in this complex structured, hydrogen bonded crystal goes beyond the main task of our research.

It is worth noting that the \textit{ab initio} lattice dynamics study of two other representatives of the KH$_2$PO$_4$ type family, i.e. CsH$_2$PO$_4$ and PbHPO$_4$, revealed phonon spectra stable in the $\Gamma$ point \cite{CDP,LHP}. The reason of such a lattice stability may be the one-dimensionality of the hydrogen bonds network, which is inherent for CsH$_2$PO$_4$ and PbHPO$_4$, and consequently the lower level of lattice anharmonicity in these two crystals. The lattice instability of KH$_2$PO$_4$ with large imaginary values of vibrational frequencies may indirectly imply the exclusively significant impact of hydrogen anharmonicity in lattice dynamics of KH$_2$PO$_4$ crystal even when compared with other members of this wide family compounds.

\begin{table*}[]
\caption{\small Comparison between the phonon frequencies of bulk KH$_2$PO$_4$ calculated at the $\Gamma$ point (sp. gr. I$\bar{4}$2d) and the experimental frequencies detected in SiO$_2$:KH$_2$PO$_4$ nanocomposite. RS corresponds to Raman scattering. All frequencies are indicated in cm$^{-1}$. $\nu_1$, $\nu_2$, $\nu_3$, $\nu_4$ refer the PO$_4$ internal modes and H corresponds to hydrogen vibrations. All modes of frequencies below $\sim$220 cm$^{-1}$ are predominantly external lattice vibrations. TO and LO correspond to transverse and longitudinal optic modes, respectively.}
 \label{Compar}
\footnotesize
\begin{tabular} {rcrcrccrccr}
\\ \hline
\multicolumn{2}{c}{4$A_1$} & \multicolumn{2}{c}{6$B_1$}& \multicolumn{3}{c}{7$B_2$}&
\multicolumn{3}{c}{13E}&5A$_2$\\ \hline
calcul.&RS  & calcul.&RS &\multicolumn{2}{c}{calcul.} &RS &\multicolumn{2}{c}{calcul.}&RS&calcul.
\\&&&&     TO&LO&&TO&LO&&
\\ \hline
 &&&&\multicolumn{2}{c}{\textit{539} Im}&&\multicolumn{2}{c}{\textit{514} Im}&
\\ &&&&\multicolumn{2}{c}{acoust.}&&\multicolumn{2}{c}{acoust.}&
\\$\nu$+H 310&356&125&&184 &323&190&106& 116&95&230
\\$\nu_4$+H 562&&176&&$\nu_2$+H 389& 397&391&134& 138&&$\nu_2$+H 359
\\ $\nu_1$+H 926&912&$\nu_{2,4}$+H 481&463&$\nu_4$+H 559& 576&&142& 217&153&$\nu_1$+H 851
\\ H 1346&&$\nu$+H 693&&$\nu_3$+H 1225& 1225&&222& 347&&H 1202
\\&&$\nu_3$+H 1041&&H 1280& 1309&&$\nu_4$+H 400& 495&&H 1358
\\&&H 1355&&&&&$\nu_4$+H 530& 677&
\\&&&&&&&$\nu$+H 686&926&
\\&&&&&&&$\nu_3$+H 1051&1077&994
\\&&&&&&&$\nu_3$+H 1191&1295&
\\&&&&&&&H 1300&1373&
\\&&&&&&&H 1380&1508&
\\\hline
\end{tabular}
\end{table*}

Analyzing the calculated \textit{eigen}-vectors, one may unveil the origin of the phonon modes on the atomic level. It is instructive to compare the type of calculated phonon modes (see Table \ref{Compar}) with the demands of the group theory  analysis (Table \ref{Classif}) valid for an ideal KH$_2$PO$_4$ crystal of I$\bar{4}$2d symmetry. It can be seen that only the highest-frequency modes of all symmetries meet the group theory requirements being purely hydrogen H vibrations. All other calculated normal modes are of mixed type with contributions from hydrogen vibrations. Although, the type of PO$_4$ groups internal vibrations, bending $\nu_2$, $\nu_4$ and stretching $\nu_1$, $\nu_3$, is predominantly preserved. There are only a few exceptions. Instead of the $\nu_2$ mode of A$_1$ symmetry the \textit{ab inito} calculation gives the $\nu_4$ mode at 562 cm$^{-1}$. There are also two internal modes near 686 cm$^{-1}$ (E) and 693 cm$^{-1}$ (B$_1$) which can not be clearly classified as either stretching or bending vibrations. A similar uncertainty in the determination of the phonon mode type is found for the calculated mode at 310 cm$^{-1}$ (A$_1$ type) which is placed in the intermediate region between the low-frequency external and the mid-frequency internal bending PO$_4$ group vibrations. Strong experimental lines detected in the Raman spectrum both in our nanocomposite material at 356 cm$^{-1}$ and in the bulk KH$_2$PO$_4$ at 354 cm$^{-1}$ (Ref. \cite{Kim}) evidence a significant phonon mode mixing in the KH$_2$PO$_4$ crystal. Note that according to the \textit{eigen}-vector analysis all external normal modes of frequencies below $\sim$220 cm$^{-1}$, translational or librational vibrations, have certain contributions of hydrogen vibrations as well. This also concerns the two unstable modes, \textit{Im} 539 and \textit{Im} 514 cm$^{-1}$, which are schematically presented in Figure \ref{eigen}.  As seen in Figure \ref{eigen}(a), the atom displacements corresponding to B$_2$ unstable mode (\textit{Im} 539 cm$^{-1}$) contain the shifts of P and K atoms along $z$ axis in opposite directions, deformation of PO$_4$ groups and the proton ordering near the oxygen atoms placed on the top of every PO$_4$ tetrahedra. The largest displacements are inherent for H and P atoms, whereas the displacements of K and O atoms are one order of magnitude smaller.

The proton distribution realized due to the condensation of B$_2$ mode near Brillouin zone center, i.e. only two hydrogen atoms near every PO$_4$ group, corresponds to lowest energy of the 16 PO$_4$ group configurations in conformity with Slater-Takagi model \cite{Blinc1987}. According to $eigen$-vectors of two-fold degenerated E phonon mode (see Figure \ref{eigen}(b,c)) each pair of H atoms shift in the same direction along the hydrogen bonds directed along $x$ or $y$ axes. However, the values of displacements differ almost two orders magnitude for each pair of protons (bigger and smaller arrows near hydrogen atoms in Figure \ref{eigen}(b,c)) realizing thus the proton configuration with energy $w$ according to Slater-Takagi model \cite{Blinc1987} (one proton is placed near PO$_4$ group), which is bigger then the proton energy of PO$_4$ group configuration depicted in Figure \ref{eigen}(a). Most probably, since the \textit{Im} 539 cm$^{-1}$ B$_2$ mode has larger imaginary frequency comparing with \textit{Im} 514 cm$^{-1}$ E modes, the loss of lattice stability regarding to the B$_2$ mode evokes the structural transition to new phase. However, the real atom displacements occurred at I$\bar{4}$2d to Fdd2 structural transition \cite{Itoh1998,Miyoshi2011} are rather complex and cannot be entirely described by $eigen$-vector of unstable B$_2$ optic mode but is a result of many-phonon anharmonic interaction. These observation confirms the exceptionally important role of the proton subsystem in the lattice dynamics of KH$_2$PO$_4$ crystal and especially in the ferroelectric phase transition mechanism which is of mixed "order-disorder" and "displacive" type.

Now one may look closer at the origin of line at 522 cm$^{-1}$ which was not resolved in Raman spectrum of our nanosized KH$_2$PO$_4$ sample (Figure \ref{Raman}(b)) but was observed in the bulk crystal (Figure \ref{Raman}(c))\cite{Kim}. According to Table \ref{Compar}, the nearest calculated frequencies to the 522 cm$^{-1}$ one are the two-fold degenerated 530 cm$^{-1}$ (E symmetry) and the modes at 559 cm$^{-1}$ (B$_2$) and 562 cm$^{-1}$ (A$_1$). As follows from the $eigen$-vector analysis, all these calculated modes correspond to bending $\nu_4$ dynamic deformations of PO$_4$ accompanied by hydrogen vibrations. Taking into consideration the observation described in the previous section, that the relative Raman intensities of nanocomposite sample differ mainly for low-frequency modes which correspond to external vibrations of the PO$_4$ groups one may conclude that the crystal field in nanosized pores of the SiO$_2$ host impedes to some extent the PO$_4$ vibrations.

\section{Conclusions}
We successfully embedded by a solution-based crystallization approach nanosized KH$_2$PO$_4$ crystals with an average size of $\sim$ 20~nm inside the nanopores of a monolithic SiO$_2$ matrix. The X-ray diffraction pattern of the powdered SiO$_2$:KH$_2$PO$_4$ sample displays the prominent Bragg peaks of the bulk KH$_2$PO$_4$ material. The Raman spectrum of SiO$_2$:KH$_2$PO$_4$ composite exhibits the main features of the corresponding bulk KH$_2$PO$_4$ crystal. The phonon frequencies of all vibrational regions inherent for bulk KH$_2$PO$_4$, i.e. low-frequency external, mid-frequency bending and high-frequency stretching PO$_4$ groups, were resolved in SiO$_2$:KH$_2$PO$_4$ Raman spectrum. Only one bending $\nu_4$ internal mode of PO$_4$ groups was not detected in our spectrum probably owing to small intensity of this mode in confined nanosized KH$_2$PO$_4$.  We observed some intensity line redistribution in the nanocomposite sample, especially in its low-frequency range, which is not typical for the Raman spectrum of bulk KH$_2$PO$_4$. Based on our experimental findings one can conclude that within the silica pores of $\sim$12 nm diameter ($\sim$16 unit cells of KH$_2$PO$_4$ crystal, a=b=0.74034 nm \cite{Bacon}), the confinement effect has a marginal impact on the KH$_2$PO$_4$ crystal structure, rather the main structural features of I$\bar{4}$2d symmetry is preserved. 

To validate both the Raman spectra of the confined and unconfined KH$_2$PO$_4$ crystals an \textit{ab-inito} lattice dynamics calculation was performed. It unveils a lattice instability of the crystal structure described by the I$\bar{4}$2d space group. This symmetry corresponds to time and space averaged proton arrangement in O-H$\cdot\cdot\cdot$O hydrogen bonds which may be a reason of the vibrational instability in the KH$_2$PO$_4$ crystal. Calculated \textit{eigen}-vectors reveal a mixed type of the majority of phonon modes with significant hydrogen contribution to most of them. In our opinion, this evidences the extremely important impact of the hydrogen subsystem, both in the lattice dynamics and in the structural phase transition mechanism of KH$_2$PO$_4$ crystals.

The successful design of the nanocomposite and the preserved structural and dynamical properties of the KH$_2$PO$_4$ crystals in pore space compared to the bulk crystals evidence that the synthesis approach presented here is particularly promising to integrate optical functionalities in nanoporous media. It is also a fine example, how mesoscale self-organized porosity in solids can be combined with crystallization and thus self-organization processes at molecular scales to fabricate bulk monolithic nanostructured materials with mechanical rigidity. This is still a particular challenge for the embedding of functional nanocomposites in macroscale devices \cite{Begley2019}.

\section{Acknowledgement}
Ya. Shchur was supported by project No.6541030 of National Academy of Sciences of Ukraine. A.V. Kityk acknowledges a support from resources for science in years 2018-2022 granted for the realization of international co-financed project Nr W13/H2020/2018 (Dec. MNiSW 3871/H2020/2018/2). The presented results are part of a project that has received funding from the European Union Horizon 2020 research and innovation programme under the Marie Sklodowska-Curie grant agreement no. 778156. PH thanks for support by the Deutsche Forschungsgemeinschaft (DFG, German Research Foundation) Projektnummer 192346071, SFB 986 ''Tailor-Made Multi-Scale Materials Systems'', as well as by the Centre for Integrated Multiscale Materials Systems CIMMS, funded by Hamburg science authority.
Calculations have been carried out using resources provided by Wroclaw Centre for Networking and Supercomputing (http://wcss.pl), grant No. 160.


\end{document}